\documentclass[prd,nofootinbib,preprint,epsfig]{revtex4}
\usepackage{amssymb}
\def\lsim{\mathrel{\rlap{\lower4pt\hbox{\hskip1pt$\sim$}}
    \raise1pt\hbox{$<$}}}      
\def\gsim{\mathrel{\rlap{\lower4pt\hbox{\hskip1pt$\sim$}}
    \raise1pt\hbox{$>$}}}      

\begin{document}
\title{ The  seesaw mechanism at TeV scale in the 3-3-1 model with  right-handed neutrinos }
\author{D. Cogollo\footnote{E-mail: diegocogollo@fisica.ufpb.br}, H. Diniz\footnote{E-mail: hermes@fisica.ufpb.br}, C. A. de S. Pires\footnote{E-mail: cpires@fisica.ufpb.br}, P. S. Rodrigues da Silva\footnote{E-mail: psilva@fisica.ufpb.br}}
	\affiliation{Departamento de F\'{\i}sica, Universidade Federal da
	Para\'{\i}ba, Caixa Postal 5008, 58059-970, Jo\~ao Pessoa - PB,
	Brazil.}


\begin{abstract}
\vspace*{0.5cm}
We implement the seesaw mechanism in the 3-3-1 model with right-handed neutrinos. This is accomplished by the introduction of  a scalar sextet into the model and the spontaneous violation of the lepton number. We identify the Majoron as a singlet under $SU_L(2)\otimes U_Y(1)$ symmetry, which makes it safe under the current bounds imposed by electroweak data. The main result of this work is that the seesaw mechanism works already at TeV scale with the outcome that the right-handed neutrino masses lie in the electroweak scale, in the range from MeV to tens of GeV. This window provides a great opportunity to test their appearance at current detectors, though when we contrast our results with some previous analysis concerning detection sensitivity at LHC, we conclude that further work is needed in order to validate this search.
 
\noindent
\end{abstract}
\maketitle

%
\section{Introduction}

The standard model (SM) of electroweak interactions does not predict masses for the neutrinos and any evidence that neutrinos are massive particles is a signal of new physics. Present
solar~\cite{solar}, atmospheric~\cite{atm}, reactor~\cite{reactor} and accelerator~\cite{acel} data on neutrino oscillation experiments provide convincing evidence that neutrinos are massive particles. Besides, these experiments revealed these neutrinos to be very light particles with masses in the eV scale. Thus, regarding neutrino masses, the challenge is twofold: to extend the SM in such a way as to generate Dirac or Majorana (or both) neutrino mass terms and to find mechanisms that explain the smallness of the observed masses. It is almost a consensus that the seesaw mechanism~\cite{seesaworiginal} provides the most elegant  explanation for the smallness of the neutrino masses. Thus any extension of the SM that accommodates this mechanism automatically turns out to be an interesting candidate to new physics beyond the SM.

The basic ingredient of the seesaw mechanism is the existence of right-handed neutrinos along with the assumption  that lepton number must be violated  by the Majorana right-handed mass terms at an energy scale ${\cal M}$ much larger than the electroweak symmetry breaking. Such mechanism works at grand unification scale ${\cal M} \approx10^{14}$~GeV as well as TeV scale~\cite{seesawmoha,seesawattevscale}, and its most convincing signature would be the evidence for Majorana right-handed neutrinos.

Among the extensions of the SM with right-handed neutrinos in its matter spectrum, there is a specific version of the 3-3-1 model~\cite{footpp}. There are some features of this model which are worth mentioning. Namely, demanding anomaly cancellation and QCD asymptotic freedom, three family of fermions are required to keep theoretical consistency, which somehow explains the problem of family replication; the model provides the correct pattern of electric charge quantization~\cite{ecq}; Peccei-Quinn symmetry~\cite{pq} can naturally be incorporated in the 3-3-1 model, addressing the traditional solution to the strong CP problem~\cite{pal}; it contains at least one weakly interacting massive particle, a scalar bilepton, which is the most likely candidate to explain the cold dark matter problem~\cite{wimp}.
We think that these and several other aspects of the model make it an interesting proposal for new physics beyond the SM model.

In this work we develop the neutrino sector of the 3-3-1 model with right-handed neutrinos~\cite{footpp} with the intention  of implementing the  seesaw mechanism in it. For this we add a scalar sextet to its original scalar content.  This sextet  contains a neutral component  that carries two units of lepton number and when it develops a vacuum expectation value (VEV), lepton number symmetry is spontaneously broken and  the right-handed neutrinos develop Dirac and Majorana mass terms.  As we will see, this will engender the seesaw mechanism.  We argue that the spontaneous breaking of the lepton number in this model occurs in conjunction with the spontaneous breaking of the 3-3-1 symmetry. This is an interesting possibility since there are two natural mass scales in this model, the highest being the 3-3-1 breaking scale, around TeV, and the electroweak one at 246~GeV. In this way, both symmetries, leptonic and 3-3-1 gauge symmetry, are broken at the same energy scale around TeV.  This means that the seesaw mechanism in this case will work at TeV scale. Moreover, the model is built in such a way that  all neutrino mass terms originate from the same Yukawa interaction. As we will see, this will imply right-handed neutrinos with mass at electroweak  scale, appropriate for direct search at the LHC.

This work is organized as follows. In Sec.~(\ref{sec2}) we present
the particle content of the model. Next, in 
Sec.~(\ref{sec3}) we proceed with the spontaneous breaking of the
lepton number and identify the Majoron. In Sec.~(\ref{sec4})
we implement the seesaw mechanism.  Finally, in
Sec.~(\ref{sec5}), we present our conclusions.

\section{The model }
\label{sec2}

The model we develop here is a version of the gauge models allowed by the gauge symmetry $SU(3)_C \otimes SU(3)_L\otimes U(1)_N$ (3-3-1) where right-handed neutrinos appear in its matter content as part of the lepton triplets~\cite{footpp}. People refer to this model as the 3-3-1 model with right-handed neutrinos(331$\nu_R$). We will concentrate on the
leptonic and the scalar sectors, the only ones relevant for our purpose. In regard to the other sectors,  we just mention their matter content. For example,  the model possesses nine gauge bosons~\cite{long} where four of them are  the standard gauge bosons $A_\mu$, $Z^0_\mu$ and $W^{\pm}_\mu$ and the remaining ones are called $Z^{\prime}_\mu$, $V^{\pm}$, $U^0$  and $U^{0 \dagger}$ and the last four carry two units of lepton number, the so called  bileptons~\cite{majoron}.  In the quark sector, in addition to the standard quarks, the model contains phenomenologically interesting new quarks~\cite{footpp,long} that present the uncommon feature of carrying baryon as well as lepton number, i.e., they are leptoquarks~\cite{majoron}.

In what concerns the leptons, they are arranged in triplets and singlets~\cite{footpp} as follows,
\begin{eqnarray}
f_{aL} = \left (
\begin{array}{c}
\nu_a \\
e_a \\
\nu^{c}_a
\end{array}
\right )_L\sim(1\,,\,3\,,\,-1/3)\,,\,\,\,e_{aR}\,\sim(1,1,-1),
 \end{eqnarray}
with $a=1,2,3$ representing the three known generations. We are
indicating the transformation under 3-3-1 symmetry after the similarity
sign, ``$\sim$''.  Notice that the third component of the lepton triplet is the right-handed neutrino.

Also, we form the scalar sector of the model by adding the triplets, 
\begin{eqnarray}
\eta = \left (
\begin{array}{c}
\eta^0 \\
\eta^- \\
\eta^{\prime 0}
\end{array}
\right ),\,\rho = \left (
\begin{array}{c}
\rho^+ \\
\rho^0 \\
\rho^{\prime +}
\end{array}
\right ) ,\, \chi = \left (
\begin{array}{c}
\chi^0 \\
\chi^{-} \\
\chi^{\prime 0}
\end{array}
\right ) , \label{scalartripletcontent} 
\end{eqnarray}
which are sufficient to generate the correct masses of all quarks, charged leptons and gauge bosons~\cite{footpp,long}. Note that  $\eta$ and $\chi$ both transform as $(1\,,\,3\,,\,-1/3)$
while $\rho$ transforms as $(1\,,\,3\,,\,2/3)$. 

In order to work with a minimal scenario, we impose a  set of discrete symmetries to the lagrangian of the model. As this work deals with leptons and scalars, we fix the discrete symmetries for these particles only, which are chosen as $(e_{aR},\,\rho,\, \chi) \rightarrow -(e_{aR},\,\rho,\, \chi)$. With this, only the charged leptons develop masses at tree level through the Yukawa interaction $g\bar f_{L} \rho e_{R}$. 

In order to realize the seesaw mechanism in this model at TeV scale  we need to add a scalar sextet to the original scalar content~\cite{scalarsextet},
\begin{eqnarray}
S=\frac{1}{\sqrt{2}}\left(\begin{array}{ccc}
\Delta^{0} & \Delta^{-} & \Phi^{0} \\
\Delta^{-} & \Delta^{--} & \Phi^{-} \\
\Phi^{0} & \Phi^{-} & \sigma^{0}
\end{array}\right ).
\label{scalarsextet} 
\end{eqnarray}

A remarkable fact about this sextet is that  after the  $SU(3)_C \otimes SU(3)_L \otimes U(1)_N$ symmetry breaking to the $SU(3)_C \otimes SU(2)_L \otimes U(1)_Y$ one, the sextet splits into a triplet plus a doublet and a singlet of scalars,
\begin{eqnarray}
S \rightarrow \Delta_{({\bf 1},{\bf 3},Y_\Delta)} +\Phi_{{({\bf 1},{\bf 
2},Y_{\Phi})}} + \sigma^0_{({\bf 1},{\bf 1},Y_{\sigma^0})},
\end{eqnarray}
where $Y$ are the hypercharges of the respective multiplets, with $Y_\Delta=-2$, $Y_{\Phi}=-1$ and $Y_{\sigma^0}=0$, and 
\begin{eqnarray}
\Delta=\frac{1}{\sqrt{2}}\left(\begin{array}{cc}
\Delta^{0} & \Delta^{-}  \\
\Delta^{-} & \Delta^{--} 
\end{array}\right )\,\,,\,\,\Phi=\frac{1}{\sqrt{2}}\left(\begin{array}{c}
\Phi^{0}   \\
 \Phi^{-} 
\end{array}\right )\,\,,\,\,\frac{\sigma^0}{\sqrt{2}}.
\label{breakofS} 
\end{eqnarray}
The Yukawa interaction we can construct with the sextet $S$  and the leptonic triplet $f$ is $G_{ab}\bar{f^C_{a_L}}Sf_{bL}$ which separates into the following ones, 
\begin{eqnarray}
\bar{f^C_{L}}Sf_{L}  \rightarrow	\bar L^C \Delta L + \bar L \Phi \nu_R + \bar \nu^C_R \sigma_0 \nu_R,
\label{breakoftheyukawainteractions}
\end{eqnarray}
when the 3-3-1 symmetry breaks to the $SU(3)_c\times SU(2)_L\times U(1)_Y$ symmetry.
In the above expression $L$ represents the usual SM doublet of left-handed leptons.
It is easy to see that when $\Delta^0$, $\Phi^0$  and $\sigma^0$ develop nonzero VEV, then Dirac and Majorana mass terms are generated for the neutrinos. Hence, as we can see, the 331$\nu_R$ with the sextet  has all the ingredients to implement the  seesaw mechanism. Before we go into the details of the mechanism, we must remember that in the seesaw mechanism we are going to develop here, lepton number is broken spontaneously.  In this case we must take care with the sort of Majoron that will arise from the corresponding lepton number violation.

\section{The Majoron}
\label{sec3}
When lepton number is spontaneously broken a pseudo-Goldstone boson called  Majoron must arise~\cite{majoronmoha}. Majoron is a dangerous particle, thus in order to check if we have it under control we need to develop the potential of the model and obtain the Majoron profile. Note that the three triplets together with the sextet involve eight neutral scalars. However it is required that only five of them develop VEV to trigger  the spontaneous breaking of both, the 3-3-1 symmetry and the lepton number one and then generate the correct masses for all particles of the model, including the Dirac and Majorana neutrino mass terms that give rise to the seesaw mechanism. The appropriate choice of the five neutral scalars that will develop VEV and do this job is,
\begin{equation}
\chi^{\prime 0}, \rho^{0}, \eta^{0}, \Phi^{0}, \sigma^{0} \rightarrow \frac{1}{\sqrt{2}}(v_{\chi^{\prime}, \rho, \eta, \Phi, \sigma}+R_{\chi^{\prime}, \rho, \eta, \Phi, \sigma}+iI_{\chi^{\prime}, \rho, \eta, \Phi, \sigma}).
\label{VEV}
\end{equation}

With this assignment of VEV's and the above assumed symmetries we can write the most general invariant scalar potential as,
\begin{eqnarray}
V & = & \mu_{\chi}^{2}\chi^{2}+\mu_{\eta}^{2}\eta^{2}+\mu_{\rho}^{2}\rho^{2}+\lambda_{1}\chi^{4}+\lambda_{2}\eta^{4}+\lambda_{3}\rho^{4}+\lambda_{4}(\chi^{\dagger}\chi)(\eta^{\dagger}\eta) \nonumber\\
  & + & \lambda_{5}(\chi^{\dagger}\chi)(\rho^{\dagger}\rho)+\lambda_{6}(\eta^{\dagger}\eta)(\rho^{\dagger}\rho)+\lambda_{7}(\chi^{\dagger}\eta)(\eta^{\dagger}\chi)+\lambda_{8}(\chi^{\dagger}\rho)(\rho^{\dagger}\chi) \nonumber\\
  & + & \lambda_{9}(\eta^{\dagger}\rho)(\rho^{\dagger}\eta)+(\frac{f}{\sqrt{2}}\epsilon^{ijk}\eta_{i}\rho_{j}\chi_{k}+H.C.)+\mu_{S}^{2}Tr(S^{\dagger}S) \nonumber\\
  & + & \lambda_{10}Tr(S^{\dagger}S)^{2}+\lambda_{11}[Tr(S^{\dagger}S)]^{2}+(\lambda_{12}\eta^{\dagger}\eta+\lambda_{13}\rho^{\dagger}\rho+\lambda_{14}\chi^{\dagger}\chi)Tr(S^{\dagger}S) \nonumber\\
  & + & \lambda_{15}(\epsilon^{ijk}\epsilon^{lmn}\rho_{n}\rho_{k}S_{li}S_{mj}+H.C.)+ \lambda_{16}(\rho^{\dagger}S)(S^{\dagger}\rho).
\end{eqnarray}

Considering the shift of the neutral scalars in Eq.~(\ref{VEV}), this potential provides the following set of minimum conditions,
\begin{eqnarray} &&\mu_{\chi}^{2}+\lambda_{1}v_{\chi'}^{2}+\frac{\lambda_{4}}{2}v_{\eta}^{2}+\frac{\lambda_{5}}{2}v_{\rho}^{2}+\frac{fv_{\eta}v_{\rho}}{2v_{\chi'}}+\lambda_{14}(\frac{v_{\Phi}^{2}}{2}+\frac{v_{\sigma}^{2}}{4})=0,\nonumber \\
&& \mu_{\eta}^{2}+\lambda_{2}v_{\eta}^{2}+\frac{\lambda_{4}}{2}v_{\chi'}^{2}+\frac{\lambda_{6}}{2}v_{\rho}^{2}+\frac{fv_{\chi'}v_{\rho}}{2v_{\eta}}+\frac{\lambda_{12}}{4}(v_{\sigma}^{2}+2v_{\Phi}^{2})=0,\nonumber\\
&& \mu_{\rho}^{2}+\lambda_{3}v_{\rho}^{2}+\frac{\lambda_{5}}{2}v_{\chi'}^{2}+\frac{\lambda_{6}}{2}v_{\eta}^{2}+\frac{fv_{\eta}v_{\chi'}}{2v_{\rho}}+
\lambda_{13}(\frac{v_{\sigma}^{2}}{4}+\frac{v_{\Phi}^{2}}{2})- \lambda_{15}v_{\Phi}^{2}=0,\nonumber \\
&& \mu_{S}^{2}+\frac{\lambda_{10}}{2}(v_{\Phi}^{2}+v_{\sigma}^{2})+\frac{\lambda_{11}}{2}(v_{\sigma}^{2}+2v_{\Phi}^{2})+\frac{\lambda_{12}}{2}v_{\eta}^{2}+\frac{\lambda_{13}}{2}v_{\rho}^{2}+\frac{\lambda_{14}}{2}v_{\chi'}^{2}-\lambda_{15}v_{\rho}^{2}=0,\nonumber\\
&& \mu_{S}^{2}+\frac{\lambda_{10}}{2}(2v_{\Phi}^{2}+v_{\sigma}^{2})+\frac{\lambda_{11}}{2}(v_{\sigma}^{2}+2v_{\Phi}^{2})+\frac{\lambda_{12}}{2}v_{\eta}^{2}+\frac{\lambda_{13}}{2}v_{\rho}^{2}+\frac{\lambda_{14}}{2}v_{\chi'}^{2}=0.
\label{minimalconditions}
\end{eqnarray}

With this we can obtain the mass matrix for the scalars. Although a rigorous study of the potential above was not done yet,  we limit our study to obtaining the profile of  the Majoron that emerges from the spontaneous breaking of the lepton number. For this we need to focus exclusively on the pseudo scalars of the model. After using the minimum conditions given above, the mass matrix for the  pseudo-scalar of the model in the basis $(I_{\chi},I_{\eta'},I_{\Delta},I_{\sigma},I_{\chi'},I_{\rho},I_{\eta},I_{\phi})^T$ is given by the following symmetric matrix
\begin{eqnarray}
\left(\begin{array}{cccccccc}
 \frac{\lambda_7}{4}v_\eta^2-\frac{fv_\eta v_\rho}{4v_\chi^{\prime}} & -\frac{\lambda_7}{4}v_\chi^{\prime} v_\eta+\frac{f}{4}v_\rho & 0&0&0&0&0&0 \\
 -\frac{\lambda_7}{4}v_\chi^{\prime} v_\eta+\frac{f}{4}v_\rho & \frac{\lambda_7}{4}v_\chi^{\prime 2}-\frac{f v_\chi^{\prime}v_\rho}{4v_\eta}&0&0&0&0&0&0  \\
 0&0  & -\frac{\lambda_{10}}{8}v_\sigma^2 & 0&0&\frac{\lambda_{10}v_\sigma v_\phi^2}{4v_\rho}&0&\frac{\lambda_{10}}{4}v_\phi v_\sigma \\
 0&0&0&0&0&0&0&0\\
 0&0&0&0&-\frac{fv_\eta v_\rho}{4v_\chi^{\prime}}&-\frac{f}{4}v_\eta &-\frac{f}{4}v_\rho &0 \\
 0&0&\frac{\lambda_{10}v_\sigma v_\phi^2}{4v_\rho} &0 & -\frac{f}{4}v_\eta &-\frac{\lambda_{10}v_\phi^4}{2v_\rho^2}-\frac{fv_\eta v_\chi^{\prime}}{4v_\rho} &-\frac{f}{4}v_\chi^{\prime} &-\frac{\lambda_{10}}{2v_\rho}v_\phi^3 \\
 0&0&0&0& -\frac{f}{4}v_\rho & -\frac{f}{4}v_\chi^{\prime} & -\frac{fv_\chi^{\prime} v_\rho}{4v_\eta} &0 \\
 0&0&\frac{\lambda_{10}}{4}v_\phi v_\sigma&0&0& -\frac{\lambda_{10}}{2v_\rho}v_\phi^3 &0& -\frac{\lambda_{10}}{2}v_\phi^2
\end{array}\right ).
\label{massscalr} 
\end{eqnarray}

Observe that in the mass matrix above $I_\sigma$ has a null eigenvalue and decouples from the other pseudo-scalars. This pseudo-scalar is the Majoron we are looking for. As  $I_\sigma$ is a singlet under the $SU(2)_L \times U(1)_Y$ symmetry, then it will not couple to the standard neutral gauge boson $Z^0$. This property guarantees that this Majoron is safe from the constraints of electroweak physics. This result implies that our Majoron manifests itself in genuine new interactions beyond SM only. It can receive a mass from radiative corrections and constitute the lightest scalar of 3-3-1 spectrum, which would make of it a possible component of cold dark matter, but it is out of our scope here to discuss such issues. Without having to worry about undesirable consequences of this pseudo-scalar we can move on and proceed to construct the seesaw mechanism in this scenario.
\section{The see-saw mechanism}
\label{sec4}
We are now ready to implement the seesaw mechanism into the model.  When $\Phi^0$  and $\sigma^0$ develop VEV, the Yukawa interactions $G_{ab}\bar{f^C_{a_L}}Sf_{bL}$ generate the following mass terms for the neutrinos in the basis $(\nu_{e_L}\,,\,\nu_{\mu_L}\,,\,\nu_{\tau_L}\,,\,\nu^C_{e_R}\,,\,\nu^C_{\mu_R}\,,\,\nu^C_{\tau_R})^T=(\nu_L\,,\,\nu^C_R)^T$,
\begin{eqnarray}
\frac{1}{2}\left( \bar{\nu^C_{L}}\,\,,\,\, \overline{\nu_{R}}\right)
\left (
\begin{array}{cc}
0 & M_{D} \\
M_{D} & M
\end{array}
\right)
\left(
\begin{array}{c}
\nu_{L} \\
\nu^C_{R}
\end{array}\right),
\label{seesawmatrix}
\end{eqnarray}
where 
\begin{eqnarray}
	M_{D}=Gv_{\Phi}\,\,\,\,\,\mbox{and}\,\,\,\,\,M=Gv_{\sigma},
	\label{MDM}
\end{eqnarray}
with $G$  being a symmetric matrix formed by the Yukawa couplings $G_{ab}$. Throughout this paper,  for simplicity, we neglect CP violation effects in the leptonic sector, which means that there are no phases in the above mass matrix and then all their elements are real.

As it is very well known, the diagonalization of the mass matrix in Eq.~(\ref{seesawmatrix}), for the case $v_{\sigma}>>v_{\Phi} $, leads to a mass relation among $M_D$  and $M$   known as the  seesaw mechanism~\cite{review}, 
\begin{eqnarray}
m_{\nu_L} \simeq -M_{D}M^{-1}M_{D}, & m_{\nu_R} \simeq M,
\label{generalmassmatrix}
\end{eqnarray}
where $m_{\nu_L}$  is a mass matrix for the left-handed neutrinos and $m_{\nu_R}$  is the mass matrix for the right-handed neutrinos.  

Now comes the main point of our work. Observe that, in terms of VEV's, $m_{\nu_L} \propto\frac{v^2_\Phi}{v_\sigma}$  and $m_{\nu_R}\propto v_\sigma$. Also, remember that $\sigma^0$ carries two units of lepton number. Hence, when $\sigma^0$ develops VEV both the lepton number as well as the 3-3-1 symmetry are spontaneously broken. This means that the energy scale of 3-3-1 symmetry breaking coincides with that of spontaneous breaking of the lepton number. Concerning 3-3-1 models the breaking is expected to occur at TeV scale, an so will the spontaneous lepton number violation. 
In other words, $v_{\sigma}$ should be around TeV. 

On the other hand, $\Phi$  does not carry lepton number. Its VEV generates, exclusively, Dirac mass terms for the neutrinos. The proposal of the seesaw mechanism is to yield neutrino masses at eV scale, which is accomplished if the ratio $\frac{v^2_\Phi}{v_\sigma}$ lies  around eV. However, as we argued above $v_\sigma$ should be around TeV, then the seesaw mechanism would be responsible for constraining the values of $v_\Phi$ to be around MeV, once $\frac{\left(\mbox{MeV}\right)^2}{\mbox{TeV}}\propto $eV.  

Besides, notice that  the masses of the left-handed and right-handed neutrinos both have their origin in the Yukawa interactions $G_{ab}\bar{f^C_{a_L}}Sf_{bL}$. This means that the Yukawa couplings $G_{ab}$ are common for both kinds of neutrinos. Then by choosing a set of values for $G_{ab}$ that generate masses for the left-handed neutrinos such as to explain the solar and atmospheric neutrino oscillation, we then infer the masses for the right-handed neutrinos.
To see this explicitly, let us substitute the matrices $M_D$  and $M$ given in Eq.~(\ref{MDM}) into the Eq.~(\ref{generalmassmatrix}). In doing this we obtain,
\begin{eqnarray}
	m_{\nu_L}=-G\frac{v_\Phi^2}{v_\sigma}\,\,\,\,\,\mbox{and}\,\,\,\,\, m_{\nu_R}=Gv_\sigma.
	\label{finalmass}
\end{eqnarray}

On choosing $v_\Phi=1$~MeV  and $v_\sigma=1$TeV, which give exactly $\frac{v_\Phi^2}{v_\sigma}=1 $eV, the range of values for the Yukawa couplings, $G_{ab}$, that leads to values for the masses of the left-handed neutrinos which explain solar and atmospheric neutrino oscillation lies  between $G_{ab}\approx10^{-2} -10^{-3}$, which implies right-handed neutrinos masses in the  MeV to GeV range. Hence we have a seesaw mechanism working at TeV scale with the right-handed neutrinos masses at the electroweak scale (sub-TeV).
This is a very striking result, since we are dealing with a realizable model with many nice features which also allows us to explain neutrino masses at a testable scale.

In order to have a taste of some numerical outcomes, let us take the following set of values for the symmetric Yukawa couplings $G_{11}= -0.003079$, $G_{12}= -0.002941$, $G_{13}= 0.002941$, $G_{22}= -0.02788$, $G_{23}= -0.02210$, $G_{33}=-0.02788$. Assuming $v_\Phi=1$~MeV  and $v_\sigma=1$TeV, we obtain, from Eq.~(\ref{generalmassmatrix}), the following entries for the mass matrix $m_{\nu_L}$
\begin{eqnarray}
m_{\nu_L}=\left(\begin{array}{ccc}
 0.003079& 0.002941& -0.002941 \\
 0.002941& 0.02788& 0.02210\\
 -0.002941& 0.02210 &0.02788
\end{array}\right )\mbox{eV}.
\label{mnuL} 
\end{eqnarray}
This mass matrix is diagonalized by the following mixing matrix
\begin{eqnarray}
U=\left(
\begin{array}{ccc}
0.809 & 0.588 & 0 \\
-0.416 & 0.572 & 0.707 \\
0.416 & -0.572 & 0.707
\end{array}
\right),
\label{mixingmatrix}
\end{eqnarray}
which implies the following mass prediction for the left-handed neutrinos
\begin{eqnarray}
	m_1\approx 5.7\times10^{-5}\mbox{eV}\,\,\,,\,\,\,m_2\approx 8.8\times10^{-3}\mbox{eV}\,\,\,,\,\,\,m_3\approx 5\times10^{-2}\mbox{eV}.
	\label{eigenvaluesm1m2m3}
\end{eqnarray}
The matrix in Eq.~(\ref{mixingmatrix}) can be parametrized in terms of mixing angles {\it a la} Cabibbo-Kobayashi-Maskawa(CKM) parameterization~\cite{CKM} and is reproduced if we take: $ \theta_{12}=36^o$, $\theta_{23}=45^o$  and $ \theta_{13}=0$. 

On the other hand, note that the eigenvalues in Eq.~(\ref{eigenvaluesm1m2m3})  imply the following values for the neutrino mass squared differences:
\begin{eqnarray}
	&&\Delta m^2_{21}=7.7\times 10^{-5}\mbox{eV}^2 ,\nonumber \\
	&& \Delta m^2_{32}=2.4\times 10^{-3}\mbox{eV}^2 .
	\label{neutrinooscillation}
\end{eqnarray}
Such  mass differences and  mixing angles explain both the solar and atmospheric neutrino oscillation according to the current data~\cite{kayser}.

Now let us focus on the masses of the right-handed neutrinos. For the same values of the Yukawa couplings and VEV's given above, we get the following entries for the right-handed neutrinos mass matrix given in Eq.~(\ref{finalmass}), 
\begin{eqnarray}
m_{\nu_R}=\left(\begin{array}{ccc}
 -3.079 & -2.941 & 2.941\\
-2.941 & -27.88 & -22.10\\
2.941& -22.10 & -27.88
\end{array}\right )\mbox{GeV}.
\label{mnuR} 
\end{eqnarray}
On diagonalizing such mass matrix we obtain the following prediction for the masses of the right-handed neutrinos
\begin{eqnarray}
&&m_{{4}}\approx 5.7\mbox{MeV}\nonumber\\
&&m_{{5}}\approx 8.8\mbox{GeV}\nonumber\\
&&m_{{6}}\approx 50\mbox{GeV}.
\label{righthandedmass}
\end{eqnarray}
Such values for the masses of the right-handed neutrinos allow this seesaw mechanism to be directly tested at future colliders.
Let us next discuss the mixing among right-handed and left-handed neutrinos and their signatures  at LHC. 

It is already well established how the mass matrix in Eq.~(\ref{seesawmatrix}) is diagonalized by a $U_{6\times 6}$ unitary mixing matrix~\cite{review}. As a consequence, the right-handed neutrinos  get mixed with the left-handed ones. Let us denote these mixing elements by $U_{l\nu_{nR}}$ where $l=e\,,\,\mu\,,\,\tau$  and $n=4,5,6$.  As a consequence of this mixing, heavy neutrinos can be tested considering their coupling with SM gauge bosons proportional to $U_{l\nu_{nR}}$.

For the case of right-handed neutrinos with mass lying in the electroweak scale, there are some constraints on their masses and mixing elements $U_{l\nu_{nR}}$ coming from  neutrinoless double beta decay($0\nu\beta\beta$)~\cite{neutrinoless} and LEP~\cite{lep} experiments. Such constraints translate into the following bounds
\begin{eqnarray}
&& 0\nu \beta \beta \rightarrow 	\sum_n{\frac{|U_{e\nu_{nR}}|^2}{M_n}}<5\times 10^{-8}\mbox{GeV}^{-1},\nonumber \\
&&\mbox{LEP}\rightarrow |U_{\mu \nu_{nR}}|^2, |U_{\tau \nu_{nR}}|^2 \lsim 10^{-4}-10^{-5}\,\,\,  \mbox{for}\,\,\, m_{\nu_{nR}} \approx 5\mbox{GeV}- 80\mbox{GeV}.
\label{neutrinolessbounds}
\end{eqnarray}

There are also some signatures of heavy neutrinos at electroweak scale being investigated considering current and future hadron collider experiments~\cite{zhang,pittau}. These searches consider lepton number conserving, and most interestingly, lepton number violation processes ($\Delta L=2$), through interaction of heavy neutrinos with SM gauge bosons. According to the Refs.~\cite{zhang,pittau} the LHC discovery of heavy neutrinos in the electroweak range (few GeV to about few hundreds of GeV) is possible for a rather large mixing, although the absence of background excess leads to bounds of the same order of magnitude as the ones obtained by LEP.

It is then necessary to check if our model  is in agreement with such bounds. Let us proceed with this checking by first noticing that the mixing matrix elements, $U_{l\nu_{nR}}$, are proportional to the ratio among the two energy scales involved in the seesaw mechanism~\cite{review}. In our case this ratio becomes,  $U_{l\nu_{nR}}\propto\frac{v_\Phi}{v_\sigma}$~\cite{mixing}. For the assumed values of the VEV's appearing in this ratio we get $U_{l \nu_{nR}}\propto 10^{-6}$. Therefore, considering this and the predicted range of right-handed neutrino  masses presented in Eq.~(\ref{righthandedmass}), it is clear that our model is in perfect agreement with the above results. 

It is opportune to call the attention to the fact that all these search analyses were based on SM gauge bosons interactions only. In the 331$\nu_R$, the right-handed neutrinos are part of a triplet together with left-handed neutrinos and charged leptons. Besides, new scalar and gauge bosons connect these leptons through new interactions in processes similar to those studied in Refs.~\cite{zhang,pittau}. 
Since some of these new interactions are not suppressed by mixing angles, the investigation of heavy neutrinos production in the context of 3-3-1$\nu_R$ should be further investigated. Of course, there will be a plethora of events involving new particles which will be as interesting to search as the heavy neutrinos, however, their discovery will only give more strength to the realization of seesaw mechanism in 3-3-1$\nu_R$ at TeV scale.

\section{Conclusions}
\label{sec5}
In this work we showed that the seesaw mechanism can be implemented into the 3-3-1 model with right-handed neutrinos if a sextet of scalars is introduced into the model. The mechanism requires that we first implement the spontaneous breaking of the lepton number. We did this in this work and argued that lepton number symmetry breaking can occur in conjunction with the spontaneous breaking of the 3-3-1 symmetry, implying that these breakings are associated to the same energy scale. Since we expect that the 3-3-1 symmetry breaks at the TeV scale, hence we are going to have a seesaw mechanism working at TeV scale. Moreover the masses of both left-handed and right-handed neutrinos arise from the same Yukawa interaction term, which means that the right-handed neutrinos will develop masses at electroweak scale. Thus, differently from the ordinary seesaw mechanism, the one we developed here   can be directly tested  at future colliders. We also analyzed the Majoron that resulted from the spontaneous breaking of the lepton number and showed that it is a singlet by the $SU(2)_L \times U(1)_Y$ symmetry, posing no threat to the model concerning current bounds from electroweak data. 

Finally, with a simplified numerical analysis, we guessed the Yukawa couplings of the model in order to recover the observed data on neutrino mass and mixing. This allowed us to predict the masses of heavy neutrinos to be in the range around 5.7~MeV to 50~GeV, with mixing angles $U_{l \nu_{nR}}\propto 10^{-6}$.
These predictions are well bellow the upper bounds obtained by LEP experiments as well as sensitivity of  LHC analyses, which consider processes involving only SM gauge bosons. In order to further constrain the heavy neutrinos of the 3-3-1$\nu_R$ model with a scalar sextet, a deeper and more complete analysis should be done, considering all new interactions with new heavy scalars and gauge bosons.
Nevertheless, the goal of this work was to show that 3-3-1$\nu_R$ model, as a good candidate for the SM extension, also provides a suitable scenario to accommodate seesaw mechanism at a testable energy scale.
%
\acknowledgments
This work was supported by Conselho Nacional de 
Desenvolvimento Cient\'{\i}fico e Tecnol\'ogico- CNPq(HD, CASP, PSRS) and by Coordena\c c\~ao de Aperfei\c coamento de Pessoal de N\'{\i}vel superior - CAPES (DC).


\end{document}